\RequirePackage{fixltx2e}					 

\documentclass[apj]{emulateapj}

\usepackage{amssymb}  						
\usepackage{amsmath}  						

\usepackage{comment}

\usepackage{hyperref}

\usepackage{graphicx}

\usepackage{braket}
\usepackage{listings}
\usepackage{placeins}

\def\nustar{{\it NuSTAR}}
\def\xmm{{\it XMM-Newton}}
\def\nustardas{\rm {\small NUSTARDAS}}

\begin{document}

\title{A Hard X-Ray Study of Ultraluminous X-ray Source NGC 5204 X-1 with \emph{NuSTAR} and \emph{XMM-Newton}}

\author{E. S. Mukherjee\altaffilmark{1},
D. J. Walton\altaffilmark{2,1},
M. Bachetti\altaffilmark{3,4},
F. A. Harrison\altaffilmark{1},
D. Barret\altaffilmark{3,4},
E. Bellm\altaffilmark{1},
S. E. Boggs\altaffilmark{5},
F. E. Christensen\altaffilmark{6},
W. W. Craig\altaffilmark{5},
A. C. Fabian\altaffilmark{7},
F. Fuerst\altaffilmark{1},
B. W. Grefenstette\altaffilmark{1},
C. J. Hailey\altaffilmark{8},
K. K. Madsen\altaffilmark{1},
M. J. Middleton\altaffilmark{7},
J. M. Miller\altaffilmark{9},
V. Rana\altaffilmark{1},
D. Stern\altaffilmark{2},
W. Zhang\altaffilmark{10} \\
}
\affil{$^{1}$ Space Radiation Laboratory, California Institute of Technology, Pasadena, CA 91125, USA \\
$^{2}$ Jet Propulsion Laboratory, California Institute of Technology, Pasadena, CA 91109, USA \\
$^{3}$ Universite de Toulouse; UPS-OMP; IRAP; Toulouse, France \\
$^{4}$ CNRS; IRAP; 9 Av. colonel Roche, BP 44346, F-31028 Toulouse cedex 4, France \\
$^{5}$ Space Sciences Laboratory, University of California, Berkeley, CA 94720, USA \\
$^{6}$ DTU Space, National Space Institute, Technical University of Denmark, Elektrovej 327, DK-2800 Lyngby, Denmark \\
$^{7}$ Institute of Astronomy, University of Cambridge, Madingley Road, Cambridge CB3 0HA, UK \\
$^{8}$ Columbia Astrophysics Laboratory, Columbia University, New York, NY 10027, USA \\
$^{9}$ Department of Astronomy, University of Michigan, 1085 S. University Ave., Ann Arbor, MI, 49109-1107, USA \\
$^{10}$ NASA Goddard Space Flight Center, Greenbelt, MD 20771, USA \\
}



\begin{abstract}
We present the results from coordinated X-ray observations of the
ultraluminous X-ray source NGC 5204 X-1 performed by
\emph{NuSTAR} and \emph{XMM-Newton} in early 2013. These
observations provide the first detection of NGC 5204 X-1 above
$10\; \mathrm{keV}$, extending the broadband coverage to
$0.3-20\; \mathrm{keV}$. The observations were carried out in
two epochs separated by approximately 10 days, and showed little
spectral variation, with an observed luminosity of $L_{\rm{X}} =
(4.95 \pm 0.11) \times 10^{39}$ erg\,s$^{-1}$. The broadband
spectrum robustly confirms the presence of a clear spectral
downturn above $10\; \mathrm{keV}$ seen in some previous
observations. This cutoff is inconsistent with the standard low/hard
state seen in Galactic black hole binaries, as would be expected from
an intermediate mass black hole accreting at significantly
sub-Eddington rates given the observed luminosity. The continuum
is apparently dominated by two optically thick thermal-like
components, potentially accompanied by a faint high energy
tail. The broadband spectrum is likely associated with an accretion
disk that differs from a standard Shakura \& Sunyaev thin disk.
\end{abstract}


\maketitle

\section{Introduction}
Ultraluminous X-ray Sources (ULXs) are off-nuclear point sources in
nearby galaxies with observed X-ray luminosities $L_X \geq 10^{39}$
erg s$^{-1}$, exceeding the Eddington limit for a $10 M_\Sun$
stellar-mass black hole (assuming isotropy). These high luminosities
could be explained by a population of $10^2-10^5 \; M_\Sun$
intermediate-mass black holes (IMBHs) accreting at sub-Eddington
rates \citep[e.g.][]{Miller04}. Alternately, these luminosities could
be explained by accretion onto stellar-remnant black accretors (potentially
reaching masses as large as $\sim$100\,$M_{\Sun}$;
\citealt{zampieri:09, belczynski:10}) if the emission is highly beamed
\citep[e.g.][]{King09}, or if these sources are able to radiate in excess
of their Eddington limit \citep[e.g.][]{Poutanen07}. For recent reviews,
see \citet{roberts:07} or \citet{feng:11}.

ULX spectra in the soft X-ray band ($\leq 10\; \rm keV$) have been
well-studied using telescopes such as \emph{XMM-Newton},
\emph{Suzaku}, and \emph{Chandra}. Their spectral shapes appear to
deviate substantially from those of known Galactic black hole binaries.
A spectral turnover at $\leq 10 \; \rm keV$ appears in most ULXs
with sufficient signal-to-noise \citep{stobbart:06, gladstone:11},
along with a soft excess usually modelled by a low temperature $\leq
0.3 \; \rm keV$ blackbody disk component \citep{Miller04}.
This disk component, if produced by a standard thin disk, could imply
the presence of an IMBH accretor. However, the temperature-luminosity
relationship of these sources does not appear to match the blackbody
emission from standard accretion disks \citep[e.g.][]{Kajava09},
although the expected scaling may be partially recovered using a fixed
absorption column between observations \citep{miller:13} or using
non-standard disk models \citep{vierdayanti:06}. The low disk
temperature can also be explained by a cool, optically thick corona
blocking the inner disk from observation \citep{gladstone:09}. This
corona would account for the continuum emission as well as the
spectral turnover. Alternate possibilities are that the soft
component originates from a strong outflow \citep[e.g.][]{Poutanen07}
or blurred line emission from highly ionized, fast-moving gas
\citep{goncalves:06}.

Until now, it has been difficult to distinguish between the spectral
models due to the limited $\sim$0.3--10.0\,keV bandpass over which
ULXs have been studied. The differences become clearer with data
above $10 \; \rm keV$ \citep[e.g.][]{walton:11a}, a region of the
spectrum that requires a focusing telescope with a broader bandpass.
The \emph{Nuclear Spectroscopic Telescope Array} (\emph{NuSTAR};
\citealt{harrison:13}), launched in June 2012, is the first orbiting
telescope with hard X-ray focusing capabilities over a large $3-79 \;
\rm keV$ bandpass. With a similar effective area to
\emph{XMM-Newton} at $\sim$6\,keV, \emph{NuSTAR} provides an
ideal complement to the current soft X-ray observatories for sensitive,
broadband studies of ULXs. Indeed, over the past two years,
\emph{NuSTAR}, \emph{XMM-Newton}, \emph{Chandra}, \emph{Swift}
and \emph{Suzaku} have undertaken joint observations of several
nearby ULXs \citep{bachetti:13, bachetti:14, rana:15, walton:13,
walton:14, walton:15a, walton:15b}.

This paper reports the results from observations of the ULX in NGC
5204, a nearby ($d=4.8$ Mpc) Magellanic-type galaxy
\citep{roberts:00}. NGC 5204 X-1 has a typical X-ray luminosity
$L_{\rm{X}} \sim 2-6\times 10^{39} \rm \; erg~s^{-1}$
\citep{roberts:04}, and is well-studied below 10 keV \citep{roberts:05,
roberts:06, vierdayanti:06}. It has been previously reported as an IMBH
candidate with long-term spectral variability \citep{feng:09}

The paper is structured as follows: in section \ref{obs} we describe the
observations and data reduction procedures, in section \ref{analysis}
we discuss the spectral analysis performed, and in section \ref{disc}
we discuss the results and summarize our conclusions.

\section{Observations}\label{obs}

In April 2013 \emph{NuSTAR} and \emph{XMM-Newton} performed two
coordinated observations of NGC 5204, approximately 10 days apart.
The \emph{NuSTAR} exposures were 96 ks and 89 ks, respectively,
and the \emph{XMM-Newton} exposures were 13 ks and 10 ks
(EPIC-pn), and 16 and 14 ks (EPIC-MOS1/2). Details of the observations
are summarized in Table \ref{observations}.

\begin{deluxetable}{llcr}
\centering
\tablecolumns{5}
\tablewidth{\columnwidth}
\tablecaption{Summary of X-ray data used in this analysis\label{observations}}
\tablehead{
\colhead{OBSID} & \colhead{Detector} & \colhead{Exposure (s)} & \colhead{Counts}}
\startdata
\cutinhead{Epoch 1 --- 2013 April 19}
0693851401 &  EPIC-pn & 13375 & 9176 \\
 & EPIC-MOS1 & 16396 & 3010 \\
 & EPIC - MOS2  & 16458 & 2846 \\
30002037002 & FPMA & 95964 & 1871 \\
 & FPMB & 95799 & 1907 \\
\cutinhead{Epoch 2 --- 2013 April 29}
0693850701 & EPIC-pn & 10415 & 6740 \\
 & EPIC-MOS1 & 14036 & 2129 \\
 & EPIC-MOS2 & 14260 & 2244 \\
30002037004 & FPMA & 88976 & 1794 \\
 & FPMB & 88854 & 1840
\enddata
\end{deluxetable}

\subsection{NuSTAR}

We reduced the \emph{NuSTAR} data for each of the two focal plane
modules (FPMA and FPMB) using standard methods with version 1.1.1
of the \emph{NuSTAR} Data Analysis Software (NuSTARDAS) and CALDB
version 20130509. We ran the \texttt{nupipeline} tool to produce
filtered event files, using all default options to remove passages through
the South Atlantic Anomaly and periods of Earth occultation, and to
clean the unfiltered event files with the standard depth correction, which
substantially reduces the internal high energy background. We then
extracted spectral products with \texttt{nuproducts}, using a
38$^{\prime\prime}$ radius extraction region around the source,
estimating the background from a $113^{\prime\prime}$ radius region
free of other point sources on the same detector as the target. The
\emph{NuSTAR} data provide a reliable detection of NGC 5204 X-1 up to
$\sim$20\,keV.

\subsection{XMM-Newton}

The {\em XMM-Newton}  data reduction was carried out with the {\em
XMM-Newton}  Science Analysis System (SAS v12.0.1). To produce
calibrated event files we used the tools {\tt epproc} and {\tt emproc} for
the pn and MOS detectors, respectively. We then filtered out periods of
high background according to the prescription in the SAS
manual.\footnote{http://xmm.esac.esa.int/} In {\tt evselect}, we used
the filters {\tt FLAG==0 \&\& PATTERN$<$4} for EPIC-pn and {\tt
FLAG==0 \&\& PATTERN$<$12} for the EPIC-MOS cameras. 
Spectra were extracted with {\tt evselect} from a 30$^{\prime\prime}$
radius region around X-1, and the background was estimated from a
blank region of radius 60$^{\prime\prime}$ on the same detector,
avoiding detector edges, bad pixels, and other visible sources. We also
avoided the detector column passing through X-1, as
recommended in the manual, to avoid the effects of charge spilling.
Ancillary responses and redistribution matrices were generaged with
{\tt arfgen} and {\tt rmfgen}, with the ELLBETA PSF correction enabled.

\begin{deluxetable}{llcr}
\centering
\tablecaption{Parameters from a power law fit of both epochs\label{powerlaw}}
\tablecolumns{4}
\tablewidth{\columnwidth}
\tablehead{
\colhead{Parameter} & \colhead{Unit} & \colhead{Epoch 1} & \colhead{Epoch 2}}
\startdata
\cutinhead{\emph{XMM-Newton}}
$n_{\rm H}$ & $10^{21} \; \rm cm^{-2}$  & $0.47\pm 0.09$ & $0.49^{+0.10}_{-0.09}$  \\
$N_{\rm pl}$ &  $10^{-4}$ & $3.36 \pm 0.12$ & $3.34 \pm 0.15$   \\
$\Gamma$ & & $2.04\pm 0.04$ & $2.06 \pm 0.04$  \\
$\rm \chi^2/d.o.f.$ & & 452/383 & 402/371 \\
\cutinhead{\emph{XMM+NuSTAR}}
$n_{\rm H}$ & $10^{21} \; \rm cm^{-2}$  &  $0.62 \pm 0.09$ & $0.65^{+0.10}_{-0.09}$  \\
$N_{\rm pl}$  &  $10^{-4}$ &  $3.54 \pm 0.12$ & $3.55^{+0.16}_{-0.15}$  \\
$\Gamma$ & &  $2.12 \pm 0.04$ & $2.15 \pm 0.04$ \\
$\rm \chi^2/d.o.f.$ & & 591/451 & 538/436
\enddata
\end{deluxetable}

\section{Spectral Analysis}\label{analysis}

\subsection{General Procedure}

The spectral analysis for this work was conducted using the Interactive
Spectral Interpretation System (ISIS) \citep{houck:00}. ISIS was chosen
over the more widely used XSPEC \citep{arnaud:96} for ease of
programmability and its transparent use of parallelized fitting and error
bar searches while also including all XSPEC models and tables. 

We modeled the neutral absorption column using \texttt{tbnew}, a
newer version of \texttt{tbabs} \citep{wilms:00}, with the absorption
cross-sections of \citet{verner:96} and appropriate solar abundances.
Cross-calibration between the various detectors was addressed using a
multiplicative constant fixed to 1 for EPIC-pn and allowed to float
otherwise; the calibrations of \xmm\ and \nustar\ are known to generally
show a good agreement (\citealt{madsen:15}). We performed fitting using
$\chi^2$ minimization and quote errors as $90\%$ confidence intervals
unless stated otherwise. During our spectral analysis, all datasets
were grouped to a minimum of 30 counts per bin to facilitate the use of
$\chi^{2}$ statistics.

Fitting the \emph{XMM-Newton} and broadband spectra independently
with a simple power law indicates low variability between the epochs,
summarized in Table \ref{powerlaw}. The residuals for these fits
behave very similarly in both epochs. For the remainder of the analysis 
we have therefore combined the epochs using the HEASOFT tools
\texttt{addascaspec} and \texttt{addrmf} to maximize source statistics.

\subsection{Modelling}\label{modeling}

\begin{figure}[t]
\hspace{-0.5cm}
\centering
\includegraphics[height=\columnwidth, angle=-90]{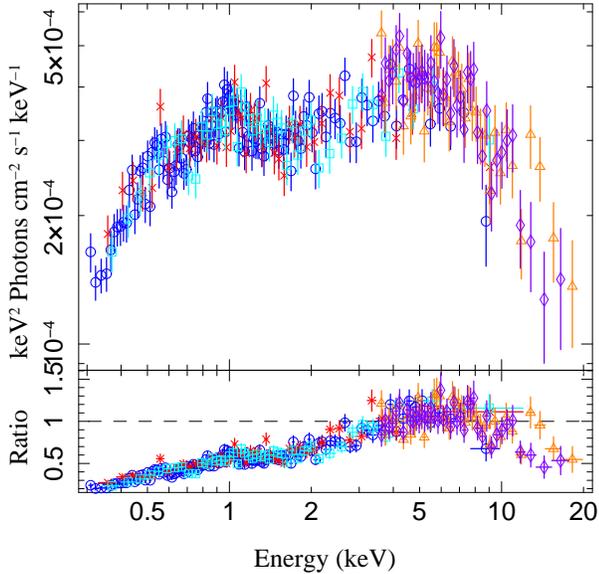}
\caption{\emph{Top panel}: Unfolded broadband spectrum of NGC 5204
X-1. EPIC-pn is plotted in blue circles and EPIC-MOS1 and MOS2 are plotted in
red stars and cyan squares, respectively. \emph{NuSTAR} FPMA and FPMB are shown
in orange triangles and purple diamonds, respectively. \emph{Bottom panel}: Data/model
ratio residuals from an unabsorbed power law evaluated over the
overlapping $3.5-10 \; \rm keV$ band, plotted over the full range. Data
were rebinned to 100 cts/bin (\emph{XMM-Newton}) and 30 cts/bin
(\emph{NuSTAR}) for presentation purposes.}
\label{noabs}
\end{figure}

We first fit the combined spectrum restricted to the $3.5-10\; \rm keV$
region of overlap between the \emph{XMM-Newton} and \emph{NuSTAR}
using a simple power law continuum and Galactic neutral absorption
column, fixed at $N_{\rm H} = 1.66 \times 10^{20} \rm \; cm^{-2}$
\citep{kalberla:05}. The result is a fit with $\chi^2_\nu = 1.18 (199/168)$,
and significant spectral curvature is evident when the model is evaluated
over the broader $0.3-20\; \rm keV$ spectral range, as shown in the
lower panel of Figure \ref{noabs}. The  $0.3-3\; \rm keV$ spectral
curvature may suggest an overall neutral absorption in excess of the
Galactic column, so for the remainder of this work, we introduce a second
absorption component intrinsic to NGC 5204. The column of this second
neutral absorption model is allowed to vary.

We fit the broadband $0.3-20\; \rm keV$ continuum using several
models frequently used to describe ULX spectra. Initially, we examine
six simple models: 1) a simple power law; 2) a power law with
exponential cutoff, XSPEC model \texttt{cutoffpl}; 3) a blackbody disk
model with a radially variable temperature index, $p$ \texttt{diskpbb}
(a ``slim disk" model with advection) \citep{mineshige:94}; 4) a simple
power law with an additional \citet{shakura:73} multicolor blackbody
disk component, \texttt{diskbb} \citep{mitsuda:84}; 5) a power law with
an exponential cutoff and an additional multicolor blackbody disk; and
6) the same, replacing the exponential cutoff with a broken power law,
XSPEC model \texttt{bknpower}. Most models we consider in this work
are fit to both the broadband \emph{NuSTAR}+\emph{XMM-Newton}
data and to the \emph{XMM-Newton} data alone, for comparison,
although we limit the majority of our description of the model fitting
to the broadband spectrum.

\begin{figure}[t]
\vspace{0.3cm}
\hspace{-0.5cm}
\centering
\includegraphics[height=\columnwidth, angle=-90]{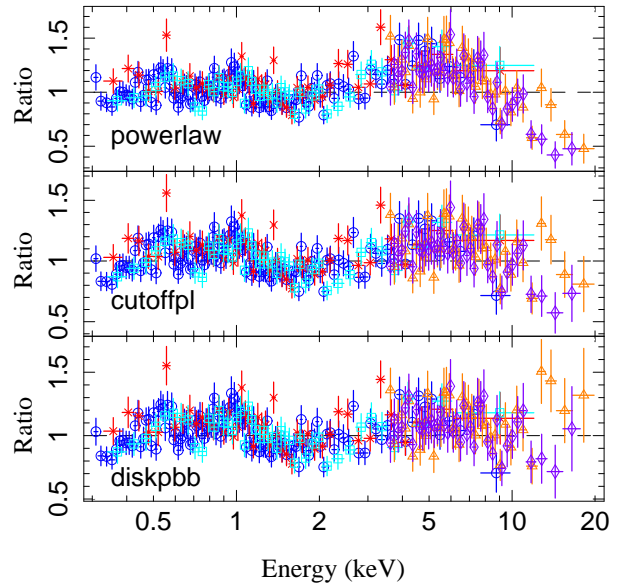}
\caption{Data/model ratios for some of the single-component models
considered here: a simple power law (top), a power law with an
exponential cutoff (middle), and an advection-dominated accretion
disk with variable temperature profile (bottom). The color scheme is
the same as Figure \ref{noabs}.}
\label{simplemodels}
\end{figure}

Best-fit parameters for each model are summarized in Table \ref{pars1}.
The simple power law model gives, as seen before, a poor fit with an
``m"-shaped structure to the data/model residuals below $\sim$10
keV and downward curvature in the $10-20\; \rm keV$ energy
range. The power law with an exponential cutoff gives an improved fit
with $\Delta \chi^2 = 107$ for one fewer degree of freedom (d.o.f.), but
the ``m"-shape is still visible in the residuals. We also attempted to
model the spectrum using just an absorbed multicolor blackbody disk,
which gave a poor fit with $\chi^2_\nu = 6.18$ (4582/741). At high
accretion rates, the expected emission likely deviates substantially from
a \citet{shakura:73} thin disk, resulting in a shallower temperature
profile (e.g. \citealt{abramowicz:88}), so we tried the \texttt{diskpbb}
model, yielding a radial temperature profile $p = 0.505 \pm 0.004$.
While the \texttt{diskpbb} is a marked improvement over the simpler
\texttt{diskbb} model, with $\chi^2_\nu = 1.21$ (890/740), once again
the ``m"-shaped residuals remain, implying the need for two continuum
components below 10\,keV. The fits with these single component models
(excluding the thin disk model, as the fit was very poor) are compared in
Figure \ref{simplemodels}.

\begin{figure}[t]
\hspace{-0.5cm}
\centering
\includegraphics[height=\columnwidth, angle=-90]{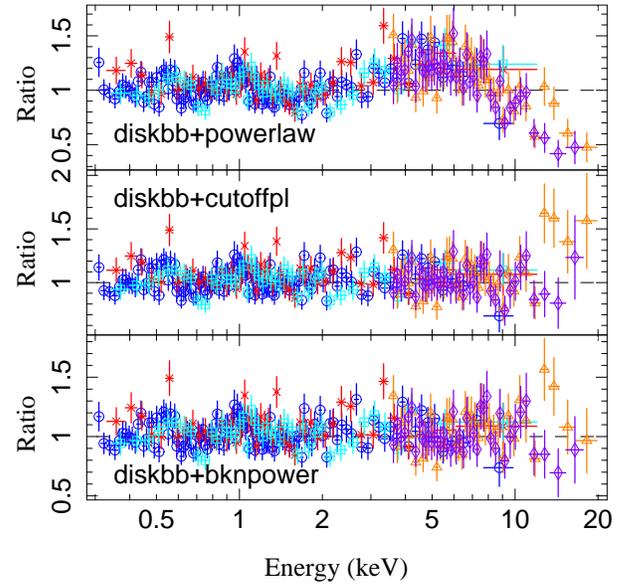}
\caption{Data/model ratios for the simple two-component continuum
models considered here: a simple powerlaw (top), a powerlaw with an
exponential cutoff (middle), and a broken powerlaw (bottom), each
combined with a \citet{shakura:73} thin disk model. The color scheme
is the same as Figure \ref{noabs}.}
\label{diskmodels}
\end{figure}

The blackbody disk component introduced in the latter three of our
simple models considerably improves the low energy excess below
$\sim 2 \; \rm keV$. Adding the disk component to the powerlaw
model provides an improvement of $\Delta \chi^2 = 12$ for 2 extra
d.o.f., but the residuals for this model still show a downturn at high
energies, so we also consider models including a disk component
and a powerlaw with an exponential cutoff, and a broken power law.
Both exponential cutoff and broken power laws provide a statistically
good fit with $\chi^2_\nu = 1.04$ (770/738 and 771/737,
respectively). The model including a blackbody disk and a broken
powerlaw is perhaps a slightly better fit, with the data distributed
more evenly about the model at high energies, although we cannot
conclusively reject the cutoff powerlaw. The fits obtained with these
two-component models are compared in Figure \ref{diskmodels}.

\begin{deluxetable}{l l c r}
\centering
\tablecaption{Best fit spectral parameters for several common simple
empirical models.\label{pars1}}
\tablecolumns{4}
\tablewidth{\columnwidth}
\tablehead{
\colhead{Parameter} & \colhead{Unit} & \colhead{\emph{XMM-Newton}} & \colhead{\emph{XMM + NuSTAR}}}
\startdata
\cutinhead{\texttt{tbnew$\times$powerlaw}}\\
$n_{\rm H}$ & $10^{21} \; \rm cm^{-2}$ & $0.49\pm 0.07$ & $0.65 \pm 0.06$\\
$N_{\rm pl}$ & $10^{-4}$ & $3.50 \pm 0.08$ & $3.70\pm 0.08$\\
$\Gamma$ & & $2.05\pm 0.03$ & $2.14\pm 0.02$\\
$\chi^2$/d.o.f. & & 716/612 & 1030/741 \\
\cutinhead{\texttt{tbnew$\times$cutoffpl}}\\
$n_{\rm H}$ & $10^{21} \; \rm cm^{-2}$ & $0.49\pm 0.07$ & $0.30^{+0.08}_{-0.07}$\\
$N_{\rm pl}$ & $10^{-4}$ & $3.51\pm 0.08$ & $3.62^{+0.08}_{-0.07}$\\
$\Gamma$ & & $2.05\pm 0.03$ & $1.84\pm 0.05$\\
$E_{\rm cut}$ & keV & $500_{-392}$ & $13.1^{+2.7}_{-198}$\\
$\chi^2$/d.o.f. & & 716/611 & 923/740 \\
\cutinhead{\texttt{tbnew$\times$diskpbb}}\\
$n_{\rm H}$ & $10^{21} \; \rm cm^{-2}$ & $0.49\pm 0.07$ & $0.37^{+0.07}_{-0.06}$ \\
$N_{\rm disk}$ & $10^{-6}$ & $4.08^{+1.13 \times 10^5}_{-1.43}$ & $83.43^{+27.32}_{-26.72}$ \\
$T_{\rm in}$ & keV & $7.22^{+0.81}_{-0.94}$ & $3.59^{+0.31}_{-0.21}$ \\
$p$ & & $0.494 \pm 0.004$ & $0.505\pm 0.004$\\
$\chi^2$/d.o.f. & & 705/611 & 890/740 \\
\cutinhead{\texttt{tbnew$\times$(diskbb+powerlaw)}}\\
$n_{\rm H}$ & $10^{21} \; \rm cm^{-2}$ & $0.65^{+0.15}_{-0.13}$ & $1.07^{+0.29}_{-0.26}$\\
$N_{\rm disk}$ & & $9.78^{+11.19}_{-4.89}$ & $141.90^{+0.32}_{-119.61}$\\
$T_{\rm in}$ & keV & $0.21 \pm 0.03$ & $0.12^{+0.03}_{-0.02}$\\
$N_{\rm pl}$ & $10^{-4}$ & $2.94^{+0.19}_{-0.20}$ & $3.74^{+0.14}_{-0.13}$\\
$\Gamma$ & & $1.89^{+0.05}_{-0.06}$ & $2.14\pm 0.02$\\
$\chi^2$/d.o.f. & & 663/610 & 1018/739 \\
\cutinhead{\texttt{tbnew$\times$(diskbb+cutoffpl)}}\\
$n_{\rm H}$ & $10^{21} \; \rm cm^{-2}$ & $0.28^{+0.16}_{-0.13}$ & $0.36^{+0.15}_{-0.13}$ \\
$N_{\rm disk}$ & & $6.44^{+3.38}_{-2.00}$ & $7.80^{+4.20}_{-2.50}$\\
$T_{\rm in}$ & keV & $0.28\pm 0.03$ & $0.25\pm 0.02$\\
$N_{\rm pl}$ & $10^{-4}$ & $2.01^{+0.41}_{-0.44}$ & $2.36^{+0.26}_{-0.27}$\\
$\Gamma$ & & $0.68^{+0.44}_{-0.53}$ & $1.09^{+0.16}_{-0.18}$\\
$E_{\rm cut}$ & keV & $3.31^{+1.78}_{-0.96}$ & $4.77^{+0.78}_{-0.64}$\\
$\chi^2$/d.o.f. & & 633/609 & 770/738 \\
\cutinhead{\texttt{tbnew$\times$(diskbb+bknpower)}}\\
$n_{\rm H}$ & $10^{21} \; \rm cm^{-2}$ & $0.49^{+0.15}_{-0.14}$ & $0.57^{+0.14}_{-0.13}$ \\
$N_{\rm disk}$ & & $6.52^{+5.00}_{-2.53}$ & $7.76^{+37.89}_{-3.40}$\\
$T_{\rm in}$ & keV & $0.25 \pm 0.03$ & $0.23 \pm 0.03$ \\
$N_{\rm pl}$ & $10^{-4}$ & $2.41^{+0.29}_{-0.33}$ & $2.68^{+0.22}_{-0.24}$\\
$\Gamma_1$ & & $1.69^{+0.10}_{-0.13}$ & $1.80^{+0.06}_{-0.07}$\\
$E_{\rm break}$ & keV & $5.07^{+0.72}_{-0.81}$ & $5.81^{+0.44}_{-0.47}$\\
$\Gamma_2$ & & $2.84^{+0.56}_{-0.45}$ & $3.00^{+0.16}_{-0.15}$\\
$\chi^2$/d.o.f. & & 633/608 & 771/737\\
\enddata
\end{deluxetable}

To explore more physical models, we investigate the possibility that
the continuum is from cool, thin disk photons Compton up-scattered
in a hot corona. We model this component using the \texttt{comptt}
model \citep{titarchuk:94}. \texttt{comptt} is an analytic
Comptonization model that assumes the seed photon spectrum
follows a Wien law with some temperature $T_{0}$. Its use allows the
temperature and the optical depth of the Comptonizing electrons to be
fit as independent parameters. This model is frequently combined with
a blackbody disk to represent a standard disk--corona accretion
geometry, with the Compton seed photon temperature linked to the
inner disk temperature. Although we also present fits with the
\texttt{comptt} model alone for completeness, we note that formally this
describes a very extreme scenario in which essentially the entire X-ray
emitting accretion disk is enshrouded by the corona, which is likely
unphysical.

\begin{figure}[t]
\hspace{-0.5cm}
\centering
\includegraphics[height=\columnwidth, angle=-90]{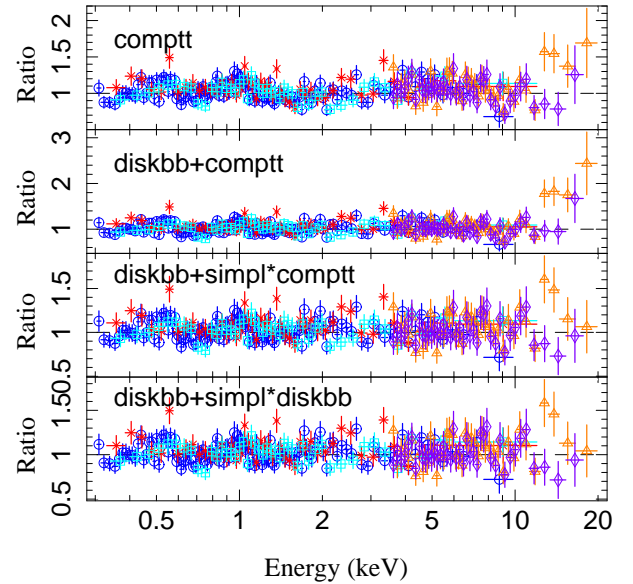}
\caption{Data/model ratios for the semi-physical models considered:
a single-component Comptonized continuum (top), thin a disk with
an additional Comptonization continuum, the same model with an
additional high-energy powerlaw tail, and the two-disk model with a
high-energy power law tail (bottom). As before, the color scheme
is the same as Figure \ref{noabs}.}
\label{compttmodels}
\end{figure}

The fit with the \texttt{comptt} model provides a reasonable statistical
fit with $\chi^2_\nu = 1.12$ (824/739), but the ``m''-shaped
residuals seen previously are again apparent below 10\,keV. Including
a blackbody disk component gives a formally acceptable fit of
$\chi^2_\nu = 1.07$ (787/738), an improvement of $\Delta \chi^2 =
37$ for one additional d.o.f., but evidence for an excess in the data
over the model remains at high energies (see Figure \ref{compttmodels}
panel 2). This is also seen with similar models in other \emph{NuSTAR}
ULX observations (e.g. \citealt{walton:13, walton:14, walton:15b}). We
address this excess with the addition of a powerlaw tail using the
\texttt{simpl} convolution model \citep{steiner:09}. This powerlaw tail
gives a fit improvement of $\Delta \chi^2 = 22$ with two additional d.o.f.
over the \texttt{diskbb+comptt} model, providing a statistically good fit
with $\chi^2_\nu = 1.04$ (765/736) and resolving the high-energy
excess seen previously. In all of these models, the \texttt{comptt}
component is cool and optically thick, resulting in a quasi-thermal
blackbody-like spectrum. We note that the assumption of linking
the Compton seed photon temperature to that of the observed disk
emission may not be valid in the case of a central optically thick corona
that obscures the inner disk (e.g. \citealt{gladstone:09}). However, we
are unable to constrain these quantities independently if they are
allowed to vary separately, and the fit remains unchanged, so we keep
them linked for convenience. Replacing \texttt{comptt} with a second
\texttt{diskbb} component in this final model results in an equally good
fit with $\chi^2_\nu = 1.04$ (766/737). Both models are illustrated in the
bottom panels of Figure \ref{compttmodels}. A full list of best-fit
parameters for these models is presented in Table \ref{pars2}.

\begin{deluxetable}{l l c r}
\centering
\tablecaption{Best fit spectral parameters for several more physical
models.\label{pars2}}
\tablecolumns{4}
\tablewidth{\columnwidth}
\tablehead{
\colhead{Parameter} & \colhead{Unit} & \colhead{\emph{XMM-Newton}} & \colhead{\emph{XMM + NuSTAR}}}
\startdata
\cutinhead{\texttt{tbnew$\times$comptt}}\\
$n_{\rm H}$ & $10^{21} \; \rm cm^{-2}$ & $0.00^{+0.08}_{}$ & $0.00^{+0.06}_{}$\\
$N_{\rm comp}$ & $10^{-4}$ & $3.50^{+1.05}_{-0.35}$ & $4.37^{+0.34}_{-0.32}$\\
$T_0$ & keV & $0.118^{+0.005}_{-0.007}$ & $0.116^{+0.005}_{-0.006}$\\
$kT$ & keV \tablenotemark{b}  & $3.25^{+496.75}_{-0.73}$ & $2.63^{+0.18}_{-0.15}$\\
$\tau_p$ & & $5.26^{+0.87}_{-5.25}$ & $6.01\pm 0.27$\\
$\chi^2$/d.o.f. & & 680/610 & 824/739\\
\cutinhead{\texttt{tbnew$\times$(diskbb+comptt)}\tablenotemark{a}}\\
$n_{\rm H}$ & $10^{21} \; \rm cm^{-2}$ & $0.29^{+0.15}_{-0.13}$ & $0.36^{+0.18}_{-0.15}$\\
$N_{\rm disk}$ & & $13.04^{+9.38}_{-5.36}$ & $24.29^{+16.40}_{-8.85}$\\
$T_{\rm in}$ & keV & $0.23^{+0.04}_{-0.03}$ & $0.19\pm 0.02$\\
$N_{\rm comp}$ & $10^{-4}$ & $2.89^{+0.40}_{-0.35}$ & $2.87^{+0.35}_{-0.30}$\\
$kT$ & keV & $1.70^{+0.30}_{-0.21}$ & $2.26^{+0.17}_{-0.14}$\\
$\tau_p$ & & $9.33^{+1.77}_{-1.37}$ & $7.07^{+0.50}_{-0.46}$\\
$\chi^2$/d.o.f. & & 633/609 & 787/738\\
\cutinhead{\texttt{tbnew$\times$(diskbb+simpl*comptt)}\tablenotemark{a}}
$n_{\rm H}$ & $10^{21} \; \rm cm^{-2}$ & \nodata & $0.30^{+0.15}_{-0.13}$ \\
$N_{\rm disk}$ & & \nodata & $9.86^{+37.73}_{-4.74}$\\
$T_{\rm in}$ & keV & \nodata & $0.26\pm 0.05$\\
$N_{\rm comp}$ & $10^{-4}$ & \nodata & $3.14^{+0.90}_{-0.40}$ \\
$kT$ & keV & \nodata & $1.30^{+0.52}_{-0.44}$\\
$\tau_p$ & & \nodata & $10.32^{+4.29}_{-2.19}$ \\
$\Gamma$ \tablenotemark{c}  & &\nodata & $3.18^{+0.73}_{-2.08}$\\
$f_{\rm scat}$\tablenotemark{d} & & \nodata & $0.61^{+0.39}_{-0.54}$\\
$\chi^2$/d.o.f. & & \nodata & 765/736 \\
\cutinhead{\texttt{tbnew$\times$(diskbb+simpl*diskbb)}}
$n_{\rm H}$ & $10^{21} \; \rm cm^{-2}$ & \nodata & $0.25^{+0.12}_{-0.11}$ \\
$N_{\rm disk_1}$ & & \nodata & $5.78^{+2.05}_{-1.51}$\\
$T_{\rm in_1}$ & keV & \nodata & $0.29 \pm 0.02$\\
$N_{\rm disk_2}$ & $10^{-2}$ & \nodata & $2.24^{+0.95}_{-1.46}$ \\
$T_{\rm in_2}$ & keV & \nodata & $1.15^{+0.45}_{-0.11}$\\
$\Gamma$ & &\nodata & $3.21^{+0.31}_{-0.73}$ \\
$f_{\rm scat}$\tablenotemark{d} & & \nodata & $1.0_{-0.70}$\\
$\chi^2$/d.o.f. & & \nodata & 766/737 \\
\enddata
\tablenotetext{a}{The Comptonization input photon temperature has
been set to the inner disk temperature.}
\tablenotetext{b}{Upper confidence bound at hard limit of $kT = 500$ for \emph{XMM-Newton} data}
\tablenotetext{c}{Lower confidence bound at hard limit of $\Gamma
= 1.1$}
\tablenotetext{d}{Upper confidence bound at hard limit of $f_{\rm
scat} = 1$}
\end{deluxetable}

\section{Discussion and Conclusions}\label{disc}

We have presented an analysis of the two coordinated \emph{NuSTAR}
and \emph{XMM-Newton} observations of the ULX NGC 5204 X-1
performed in 2013. The contribution of the \emph{NuSTAR} data has
allowed us to produce the first broadband spectrum of this source
extending above 10 keV. While NGC 5204 X-1 is a source known
previously to display aperiodic spectral variability on a time-scales
of several days (e.g. \citealt{roberts:06}), we found low variability between
the two observing epochs, separated by $\sim$10 days, and therefore
combined them to maximize count statistics for our spectral analysis.

Prior work on NGC 5204 X-1 using data from \emph{XMM-Newton}
suggested that it may be an IMBH of a few hundred solar masses,
described by a cool blackbody disk ($kT\sim 0.2$\,keV) and a hard
powerlaw tail ($\Gamma \sim 2$; \citealt{feng:09}). Statistical
evidence for curvature in the 2--10\,keV bandpass has been seen
in some previous observations (e.g. \citealt{stobbart:06}), which has
generally been used to argue in favor of the high/super-Eddington
interpretation, but not others (e.g. \citealt{gladstone:09}). However,
the limited bandpass of \textit{XMM-Newton} meant even when this
curvature was detected, it was not clear whether this represented a
true spectral cutoff (e.g. \citealt{caballero:10, walton:11a}). Indeed,
the limitations of a narrow bandpass are highlighted by the results
presented here, comparing the model fits to just the
\emph{XMM-Newton} data to those to the combined \emph{NuSTAR}
and \emph{XMM-Newton} dataset. Far more models provide an
acceptable fit to the \emph{XMM-Newton} data alone than to the
broadband spectrum.

The \emph{NuSTAR} data presented here robustly demonstrate that
the spectrum of NGC 5204 X-1 displays significant curvature above
$\sim$3\,keV, and is not powerlaw-like. This is similar to the results
obtained in other \emph{NuSTAR} ULX observations (e.g. 
\citealt{bachetti:13, rana:15, walton:13, walton:14, walton:15a,
walton:15b}). Flux calculations further demonstrate that the
proportion contribution of the hard X-ray emission from NGC 5204 X-1
is relatively small, with only $\sim$10\% of the 0.3--20.0\,keV flux
emitted above 10\,keV (see Table \ref{flux}). The $\sim$5--6\,keV
break argues against the presence of a $\sim10^3 M_\Sun$ IMBH
accreting in the ``low/hard state", assuming such black holes would
display similar accretion states to Galactic binaries, which would be
expected appear powerlaw-like in the observed bandpass
\citep{remillard:06}. Using the best fit \texttt{diskbb+simpl*comptt}
model, we calculate the observed $0.3-20 \; \mathrm{keV}$ luminosity
to be $4.95 \times 10^{39}$ erg s$^{-1}$, (see Table \ref{flux}); the flux
below $10$\,keV during this epoch is similar to that observed previously
\citep{roberts:04}. We conclude that NGC 5204 X-1 is therefore likely
a high-Eddington accretor with a more modest black hole mass.

\begin{figure*}[t]
\centering
\hspace*{-0.5cm}
\includegraphics[height=\columnwidth, angle=-90]{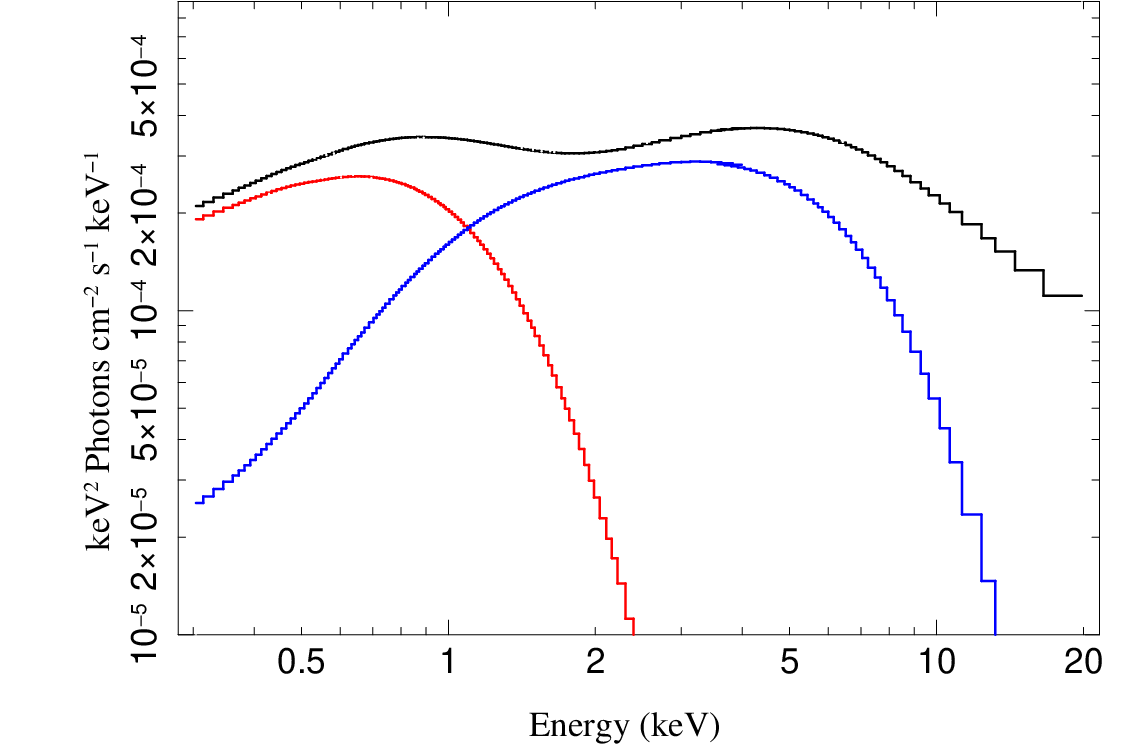}
\includegraphics[height=\columnwidth, angle=-90]{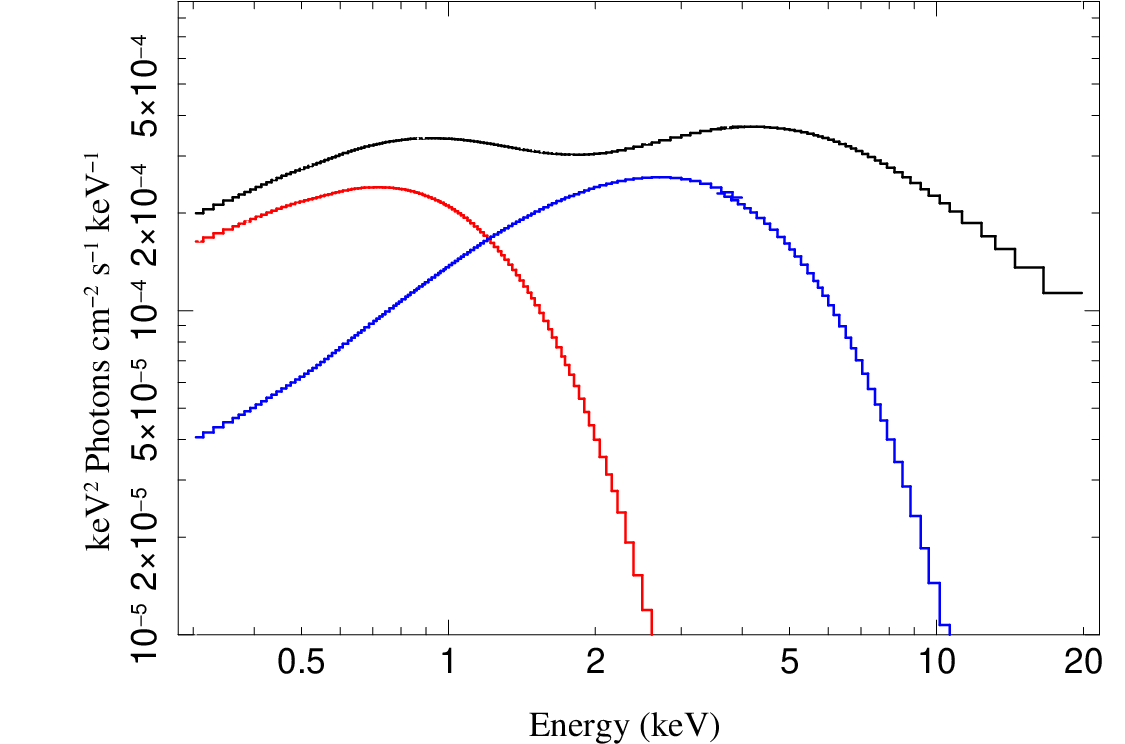}
\caption{\emph{Left panel}: the contribution of various spectral
components for the \texttt{diskbb+simpl*comptt} model. The full
model is given in black with the \texttt{diskbb} component shown
in red and the \texttt{comptt} component (before modification by
\texttt{simpl}) shown in blue. \emph{Right panel}: same as the left
panel but for the \texttt{diskbb+simpl*diskbb} model. The full
model is again given in black with the first \texttt{diskbb}
component shown in red and the second in blue.}
\label{modelcomps}
\end{figure*}

\begin{deluxetable}{l c r}
\centering
\tablecaption{Observed Fluxes for NGC 5204 X-1.\label{flux}}
\tablecolumns{3}
\tablewidth{\columnwidth}
\tablehead{
\colhead{} & \colhead{Flux ($\rm erg \; s^{-1}\; cm^{-2}$)} & \colhead{Luminosity ($\rm erg \; s^{-1}$)\tablenotemark{a}}}
\startdata
$0.3-10.0$ keV & $(1.62 \pm 0.03) \times 10^{-12}$ & $(4.47 \pm 0.08) \times 10^{39}$ \\
$10.0-20.0$ keV & $(1.74^{+0.19}_{-0.18}) \times 10^{-13}$ & $(4.80^{+0.53}_{-0.51} ) \times 10^{38}$ \\
$0.3-20.0$ keV & $(1.79 \pm 0.04) \times 10^{-12}$ & $(4.95 \pm 0.11) \times 10^{39}$
\enddata
\tablenotetext{a}{Observed luminosities (i.e. without any correction
for neutral absorption applied) calculated for a distance of $4.8$ Mpc
\citep{roberts:00}, using the \texttt{diskbb+simpl*diskbb} model}
\end{deluxetable}

The broadband spectrum in this case is best fit by a three-component
model, with two quasi-thermal components and a weak powerlaw-like
excess at high energies (Figure \ref{modelcomps}). We model the first
component using a cool, \citet{shakura:73} thin disk ($T_{\mathrm{in}}
\sim 0.25$\,keV). The hotter components can be modelled as a cool,
optically thick Comptonization model ($kT \sim 1.3$\,keV; $\tau \sim
10$), dramatically different from the hot, optically thin coronae
observed from sub-Eddington black holes (e.g, \citealt{gierlinski:99, 
miller:13b, tomsick:14, natalucci:14, brenneman:14}), or a second
multicolor blackbody disk ($T_{\mathrm{in}} \sim 1.15$\,keV).
Although poorly constrained, in both cases the best-fit photon
index of the high-energy powerlaw tail is steep ($\Gamma \sim 3$),
similar to the value of the photon index obsesrved in the steep power law
state ($\Gamma \sim 2.5$; \citealt{remillard:06}). In XSPEC syntax, the
spectrum is best described by an absorbed \texttt{diskbb+simpl*comptt}
or \texttt{diskbb+simpl*diskbb} model.

In the context of high/super-Eddington accretion, a number of physical
scenarios have been proposed for the emission components observed
from ULXs below 10\,keV. One such model invokes an optically thick
Comptonizing corona that produces the 3--10\,keV continuum and
obscures the inner portion of the accretion disk, allowing a cool disk
temperature to be observed without requiring an IMBH
\citep{gladstone:09}. Our \emph{NuSTAR} observations likely do not
favor this physical interpretation, as this model in turn seems to require
a further Comptonizing region to explain the spectrum above 10\,keV,
which appears to have some similarity with the optically-thin
Comptonization traditionally observed from Galactic binaries at high
luminosities, calling into question the initial interpretation of the
3--10\,keV continuum as the corona. Furthermore, we note that
\citet{miller:14} argue that the parameters typically obtained with
\texttt{comptt} fits to ULXs would imply very large size-scales for these
coronae, and that it is difficult to envision a physical scenario that
would result in such a large, uniformly heated region.

Alternately, \citet{middleton:11} have argued that the cool ``disk"
component could actually arise in a wind from a super-Eddington
accretion disk, and that the hotter quasi-thermal model represents
the spectrum of the inner disk itself. While we still lack
unambiguous evidence of such winds through absorption lines
\cite{walton:12, walton:13b}, this could be an effect of our viewing
angle which may not intercept the winds if the they have a roughly
equatorial geometry (e.g. \citealt{middleton:15}). In addition,
\citet{dexter:12} have recently suggested that black hole binary
accretion disks may exhibit significant inhomogeneities, resulting in
the simultaneous presence of hot and cool regions within the same
disk, and by extension an unusual disk spectrum. Such ``patchy"
disk scenarios could arise as a natural signature of photon-bubble
instabilities \citep{gammie:98} proposed to transport flux in a
super-Eddington disk \citep{miller:14}. Although there are differences
in the detailed physics, both these models associate the hotter
thermal component with emission from the accretion disk, and the
highest energy emission with optically thin Comptonization, and are
consistent with our broadband observations. With current data,
it is difficult to unambiguously associate model components with
precise physical processes, particularly with regard to the soft thermal
component; we cannot currently distinguish between a disk or
a wind origin here. Performing broadband X-ray observations of
NGC 5204 X-1 at different flux levels to probe spectral variability and
examine how the different components evolve may be the key to
distinguishing between these different models.

\section*{ACKNOWLEDGEMENTS}

The authors would like to thank the referee for their useful feedback
which helped improve the final manuscript. This research has made
use of data obtained with the \mbox{\nustar} mission, a project led by
the California Institute of Technology (Caltech), managed by the Jet
Propulsion Laboratory (JPL) and funded by NASA, and with \xmm,
an ESA science mission with instruments and contributions directly
funded by ESA Member States and NASA. We thank the \nustar\
Operations, Software and Calibration teams for support with the
execution and analysis of these observations. This research has
made use of the \nustar\ Data Analysis Software (\nustardas), jointly 
developed by the ASI Science Data Center (ASDC, Italy) and Caltech
(USA).

\bibliographystyle{/Users/dwalton/papers/apj}

\bibliography{ulx}

\end{document}